\documentclass[aps,prl, twocolumn ,superscriptaddress]{revtex4-1}

\usepackage{graphicx}
\usepackage{amsmath}

\begin{document}


\title{Spectral photonic lattices with complex long-range coupling}


\author{Bryn Bell}
\email[]{bbell@physics.usyd.edu.au}
\affiliation{Centre for Ultrahigh bandwidth Devices for Optical Systems (CUDOS), Institute of Photonics and Optical Science (IPOS), School of Physics, University of Sydney, NSW 2006, Australia}
\author{Kai Wang}
\affiliation{Nonlinear Physics Centre, Research School of Physics and Engineering, Australian National University, Canberra, ACT 2601, Australia}
\author{Alexander S. Solntsev}
\affiliation{School of Mathematical and Physical Sciences, University of Technology Sydney, Ultimo, NSW 2007 Australia}
\author{Dragomir~N.~Neshev}
\affiliation{Nonlinear Physics Centre, Research School of Physics and Engineering, Australian National University, Canberra, ACT 2601, Australia}
\author{Andrey~A.~Sukhorukov}
\affiliation{Nonlinear Physics Centre, Research School of Physics and Engineering, Australian National University, Canberra, ACT 2601, Australia}
\author{Benjamin~J.~Eggleton}
\affiliation{Centre for Ultrahigh bandwidth Devices for Optical Systems (CUDOS), Institute of Photonics and Optical Science (IPOS), School of Physics, University of Sydney, NSW 2006, Australia}


\date{\today}

\begin{abstract}
We suggest and experimentally realize a spectral photonic lattice - a  signal can hop between discrete frequency channels, driven by nonlinear interaction with stronger pump lasers. By controlling the complex envelope and frequency separations of multiple pumps, it is possible to introduce non-local hopping and to break time-reversal symmetry, which opens up new possibilities for photonic quantum simulation. As two examples, we observe a spectral quantum walk and demonstrate the discrete Talbot effect in the spectral domain, where we find novel instances containing asymmetry and periodicities not possible in spatial lattices.
\end{abstract}

\pacs{42.50.Dv, 03.67.Bg}
\maketitle
\onecolumngrid
\section{Introduction}


Discrete lattice dynamics play an important role in various branches of physics, from condensed matter to topological photonics~\cite{Lu2014}. In spatial photonic lattices, such as an array of evanescently coupled waveguides, the interactions are usually dominated by nearest-neighbor (local) coupling, with real coupling coefficients. Longer-range coupling is relevant in some electronic systems, and in photonics the addition of controlled non-local coupling provides more degrees of freedom to tailor the dispersion relation or band-structure of the lattice. Meanwhile, complex-valued coupling coefficients cause waves propagating through the lattice to accumulate a momentum-direction-dependent phase which breaks time-reversal symmetry (TRS), ie. moving from lattice site A to B imparts a different phase-shift to moving from B to A. Broken TRS is of considerable importance in topological physics, but is challenging to implement in optics, requiring for example a periodic modulation of the lattice~\cite{Rechtsman2013}. Photonic lattice systems utilizing dimensions other than space potentially allow greater freedom to realize these novel features and explore their effects; for instance, discretized spectral components of optical waves can couple to each other driven by nonlinear frequency conversion~\cite{Bersch2009}, photon-phonon interactions~\cite{Kang2009, Wolff2017}, and fast modulation of optical resonators~\cite{Yuan2016}. Temporal lattices have also been studied extensively, particularly in the contexts of quantum simulation and parity-time symmetry~\cite{Regensburger2012, Screiber2012}.

\begin{figure}[h!]
	\centering
	\includegraphics[width=.7\columnwidth]{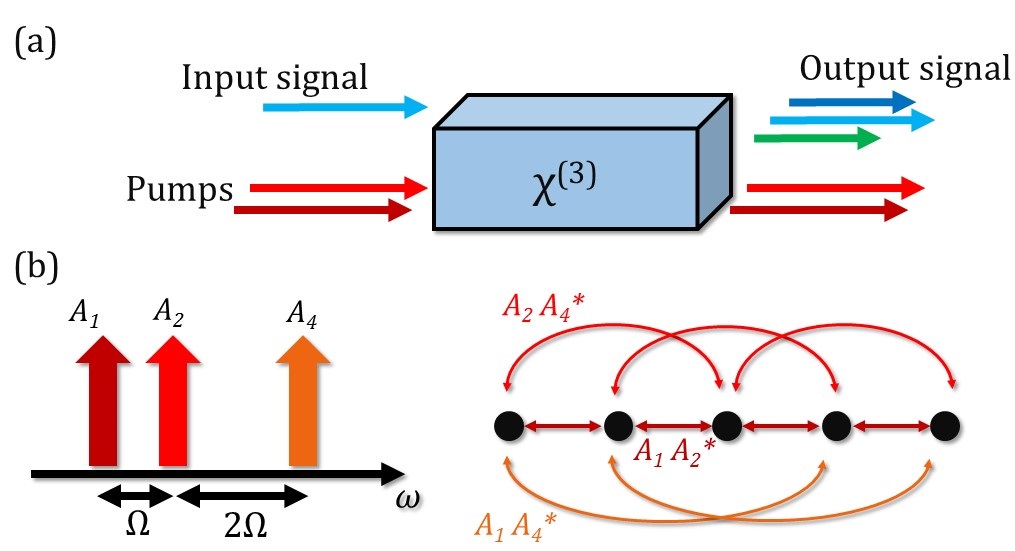}
        \vspace{-10pt}
	\caption{(a)~Four-wave-mixing in a $\chi^{(3)}$ waveguide, where pumps and signal are in multiple discrete frequencies. A pair of pumps can up- or down-shift the signal in frequency. (b)~With multiple pumps present (left), the evolution of the signal (right) is governed by multiple hopping coefficients across the lattice, each depending on the complex amplitudes of a pair of pumps. 
}\label{fig1}
 \vspace{-5pt}
\end{figure}

Here, we suggest and develop experimentally a spectrally arranged photonic lattice, where lattice sites are represented by discrete frequency channels, which are coupled together by nonlinear frequency conversion. In such a system, controllable long-range and complex coupling are made possible by shaping the spectrum of the optical pump. Our scheme promises flexibility to implement different Hamiltonians with non-trivial band-structures, particularly those breaking locality of coupling and TRS, opening up opportunities to experimentally investigate new physical effects in discrete lattices. As an example, we demonstrate a spectral analogue of the discrete Talbot effect, a self-repetitive imaging effect observed in diffractive systems with periodic input states \cite{doi:10.1080/14786443608649032, Iwanow2005}. We find two ways in which novel instances can be found which have not been possible in spatial lattices. Firstly, displaced images can be formed such that they appear to be propagating in frequency with a particular direction, arising from the breaking of TRS by complex-valued hopping. Secondly, whereas previously it was thought that an image would only occur when the input was periodic every $N=$1, 2, 3, 4, or 6 lattice sites~\cite{Iwanow2005}, we show that altering the band-structure with non-local coupling can lead to the Talbot effect occurring with other periodicities, and experimentally demonstrate the $N=5$ case. 

\section{Implementation of Spectral Lattices}

We consider a co-propagating signal and pump, whose spectra consist of a lattice of discrete channels separated in angular frequency by $\Omega$. Then in a Kerr nonlinear medium, such as optical fiber, a signal
can be up- or down-shifted by a multiple of $\Omega$ in a coherent conversion process known as four-wave mixing Bragg scattering (FWM-BS)~\cite{McKinstrie2005}, as depicted in Fig.~\ref{fig1}. Since FWM-BS is coherent and in principle noiseless, it has attracted attention in quantum optics, for manipulating the frequency of single photons~\cite{Clark2013, Bell2016, Li2016, Clemmen2016}, as well as in classical communications and all-optical switching~\cite{Zhao2017}. We assume that the signal and pump move at the same group-velocity and group-velocity dispersion can be ignored, achieved by placing them equidistant in frequency to either side of a zero dispersion wavelength. This results in a broad phase-matched bandwidth for FWM-BS. Other nonlinear processes such as parametric amplification of the signal by the pumps should be phase-mismatched to avoid introducing gain and noise into the signal.

\begin{figure}[b!]
	\centering
	\includegraphics[width=0.7\columnwidth]{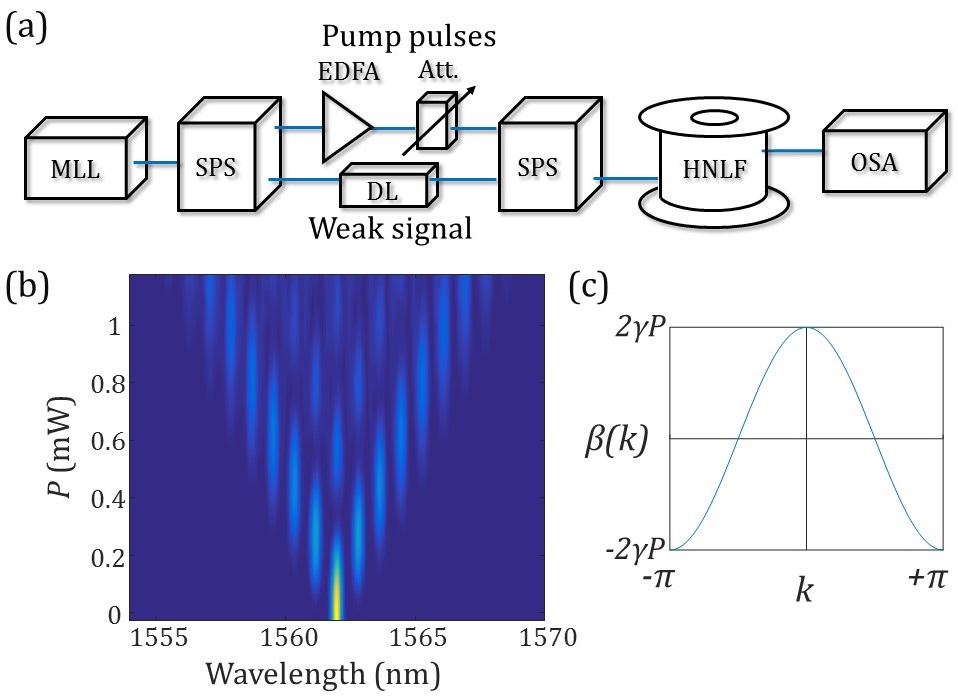}
        \vspace{-10pt}
	\caption{(a)~Experimental setup. MML: mode-locked laser; SPS: spectral pulse shaper; EDFA: erbium-doped fiber amplifier; Att.: variable attenuator; DL: tunable delay line; HNLF: highly nonlinear fiber; OSA: optical spectrum analyzer. (b)~Measurement result with two pumps, spaced 100 GHz apart around 1540nm, and a single input at 1562nm. The dynamics of the evolution are apparent as the total average pump power $P$ is varied. (c) Calculated band-structure.
}\label{fig2}
 \vspace{-5pt}
\end{figure}

The signal-envelope dynamics along propagation distance $z$ can be described by a set of amplitudes $a_n$ at frequency components $\Omega n$, which evolve according to coupled-mode equations (detailed derivation is provided in Appendix 1):
\begin{equation}\label{eq:hamiltonian}
\frac{\mathrm{d}a_n}{\mathrm{d} z}=i \sum_{j=1}^{+\infty} \left[ c_j a_{n+j}+c^\ast_j a_{n-j} \right],
\end{equation}
where $c_j = c_{-j}^\ast = 2\gamma\sum_{m} A_m(0) A_{m-j}^\ast(0)$ is the $j$th order coupling coefficient, and $A_m(z)$ are the complex amplitudes of the envelope of pump spectral components. The coupling coefficients are in general complex numbers, with a phase determined by the phases of the pumps; this can be utilized to realize a momentum-direction-dependent hopping-phase and break TRS~\cite{Yuan2016}. 
It can be understood through the effect of couplings on the band-structure of the lattice for eigenmodes $a_n(z)=\exp\{i[kn+\beta(k)z]\}$,
\begin{equation} \label{eq:model}
\beta(k)=\sum_{j=1}^{+\infty} \left[c_j e^{ijk}+c^*_j e^{-ijk}\right]=2\gamma\sum_{m\neq l} A_m(0) A^\ast_l(0) e^{i(m-l)k},
\end{equation}
The wave-number $k$ represents the reciprocal space of frequency, ie. time times lattice separation, $k=\Omega t$, which varies between $-\pi$ and $\pi$.
If the pumps do not have equal phases, in general $\beta(-k)\neq \beta (k)$, which breaks TRS. 

The experimental setup is shown in Fig.~\ref{fig2}(a). To produce multiple pump and signal frequencies which are mutually phase-stable, a single mode-locked laser with bandwidth $\sim 25$nm is filtered into the required frequencies using a spectral pulse shaper (SPS, Finisar WaveShaper 4000S). The pump frequencies are amplified then their power is controlled by a variable attenuator. A second SPS is used to recombine the signal and pumps while removing spontaneous emission from the amplifier, as well as to impart phase-shifts to the different frequencies as required. FWM-BS occurs in a 750m length of highly nonlinear fiber (HNLF). The HNLF has a zero dispersion wavelength at 1551nm, so FWM-BS is well phase-matched when the signal and pumps are evenly spaced to either side of this wavelength~\cite{Lefrancois2015}. Hence the pumps were placed around 1540nm, with a $\Omega=2\pi\times 100$~GHz frequency separation ($\sim 0.8$nm) between the channels, and the input signal was placed at 1562nm. The phase-matched bandwidth for the signal is estimated to be greater than $2\pi\times 2$THz, limited by the third order dispersion of the HNLF.

\begin{figure}[h!]
	\centering
	\includegraphics[width=0.7\columnwidth]{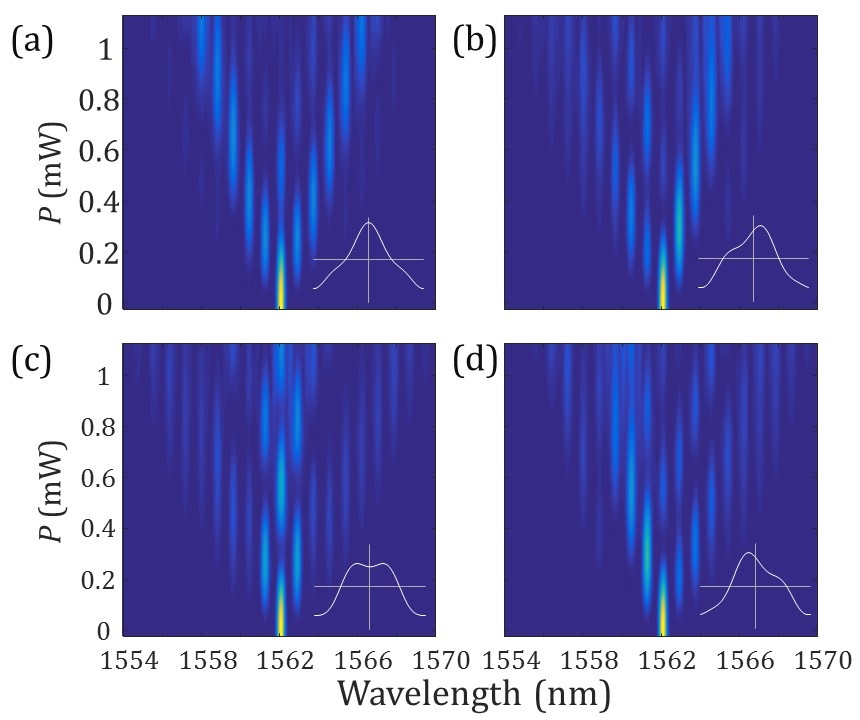}
        \vspace{-5pt}
	\caption{Measurement results with the addition of a third pump creating 2nd and 3rd order hopping $|c_3|\simeq |c_2|\simeq 0.15 |c_1|$. The phase of the new pump is set to (a) 0, (b) $\pi/2$, (c) $\pi$, and (d) $3\pi/2$. Insets to the bottom right show the calculated band-structures.
}\label{fig3}
 \vspace{-5pt}
\end{figure}

First, we realize spectral discrete diffraction in a synthetic lattice with nearest-neighbor coupling, using two equal-amplitude pumps at neighboring spectral positions, $A_1 = A_2$.
The measured spectra for this initial configuration are presented in Fig.~\ref{fig2}(b), with the vertical axis showing increasing pump power. We note that increasing the power of all pumps uniformly increases all of the $c_j$ uniformly, and so is equivalent to varying the time of the evolution, allowing the dynamics to be observed without the need to change the fiber length. 
The measurement result shows the expected features of discrete diffraction or a quantum walk, with most of the signal propagating away from the centre to the left and right of the lattice, and weak periodic revivals occurring in the central channels. In Appendix 2 we provide simulation results alongside experimental measurements showing good agreement with theory~\cite{Szameit2008}, and the fidelity between the measured and ideal spectrum remains $>95\%$ over the range of the evolution. Fig.~\ref{fig2}(c) shows the calculated band-structure for this system, which has the usual cosine-shape.

Then we implement a lattice with non-local coupling by introducing a third pump frequency, with amplitude $A_4$. 
We choose a lower power for the third component, $|A_4| \simeq 0.15 |A_{1,2}|$, creating 2nd and 3rd order hopping coefficients $c_3 \simeq c_2 \simeq 0.15 c_1$. The spectral evolution becomes highly dependent on the phase of this third pump, despite its lower power, as shown in Figs.~\ref{fig3}(a-d). For phases of 0 or $\pi$ relative to the other pumps, the evolution remains symmetric, but either signal intensity can be diverted into the propagating lobes to the left and right of the diagram [Fig.~\ref{fig3}(a)], or it can remain more localized in the central three frequency channels  [Fig.~\ref{fig3}(c)]. For phases of $\pi/2$ or for $3\pi/2$ radians [Figs.~\ref{fig3}(b,d)], the evolution becomes asymmetric; this asymmetry is also reflected in the calculated band-structure, and is connected to the breaking of TRS. This dependence on pump phase is also evidence that the different hopping terms are combining coherently, and that the overall evolution remains unitary. This would not be true if the pump frequencies were not phase-stable with respect to each other.

\section{Spectral Discrete Talbot Effect}

\begin{figure}[h!]
	\centering
	\includegraphics[width=0.7\columnwidth]{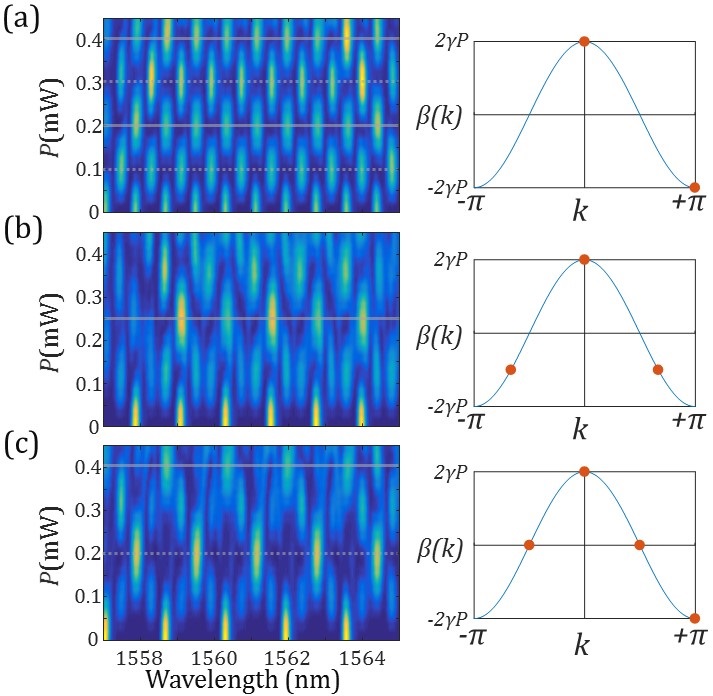}
        \vspace{-5pt}
	\caption{Experimental demonstration of spectral discrete Talbot effect, for input signals with periodicity (a)~$N=2$, (b)~$N=3$, and (c)~$N=4$. Horizontal (dashed) lines mark the positions of real (displaced) images of the input. The band-structures to the right of each measurement are marked with the positions of the non-zero $k$ components in each case (orange dots).
}\label{fig4}
 \vspace{-5pt}
\end{figure}

Next, we demonstrate a spectral discrete Talbot effect, where a periodic input pattern is recovered after a certain period of evolution. Whereas a discrete Talbot effect was previously demonstrated using optical waveguide arrays~\cite{Iwanow2005},  it was fundamentally limited only to certain periodicities due to local coupling. We show how to overcome this restriction by engineering non-local coupling, enabling self-imaging of other spectral patterns.

We consider a periodic input signal which repeats every $N$ lattice sites, and accordingly contains only a discrete set of $k$ components, $k_m = 2\pi m/N$, with $m$ an integer such that $-\pi < k_m \leq \pi$. The corresponding eigenvalues are labeled $\beta_m=\beta(k_m)$. If the ratios of the separations between $\beta_m$ are rational numbers, i.e.
\begin{equation}
\frac{\beta_i-\beta_0}{\beta_j-\beta_0}=\frac{p_{i,j}}{q},
\label{eq1}
\end{equation}
where $p_{i,j}$ and $q$ are prime numbers and $(i,j)\neq 0$, then after some period of evolution the separate $k$ components will come back into phase with one another and produce an image of the input signal. As shown in \cite{Iwanow2005}, for the case of nearest-neighbor hopping, this will only occur if $N$ is in the set $\{1,2,3,4,6\}$. 

Initially, we realize experimentally a spectral version of discrete Talbot effect for $N=2, 3, 4$, see Fig.~\ref{fig4}. Here, only two pump frequencies are used, and their separation has been decreased to $\Omega=2\pi\times 50$~GHz ($\sim 0.4$nm), to allow more periods to fit into the useful signal bandwidth.

\begin{figure}[htb]
	\centering
	\includegraphics[width=0.7\columnwidth]{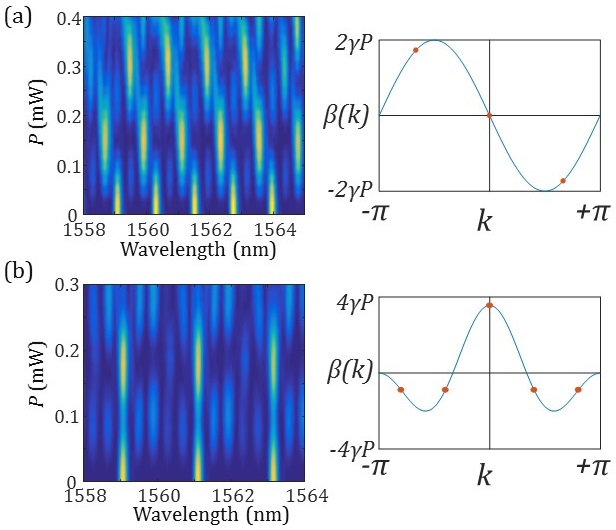}
        \vspace{-5pt}
	\caption{(a) Talbot effect combined with image shift for $N=3$. A $\pi/2$ phase-shift between the pumps creates an asymmetric propagation in which regular displaced images appear. 
    (b) Talbot effect for $N=5$, which required first and second order couplings realized with three pump frequencies.
}\label{fig5}
 \vspace{-5pt}
\end{figure}

The simplest non-trivial case is $N=2$ [see Fig.~\ref{fig4}(a)]; here, every other lattice site is initially occupied, and the state can be described by two $k$ components, $k=0, \pi$. At around 0.1mW pump power a phase-difference of $\pi$ has accumulated between these $k$ components, and the input sites are depleted of signal. At around 0.2mW, a phase-difference of $2\pi$ has accumulated,  creating an image, and then this pattern repeats. For $N=3$ [see Fig.~\ref{fig4}(b)], a higher power of around 0.25mW is required to see an image, because the allowed $k$ components are more closely spaced in $\beta(k)$ and take longer to accumulate a $2\pi$ phase difference. The imperfections in this image could be explained by errors in setting the amplitudes and phases of the input state, or a consequence of dispersion in the nonlinear fiber. Since pulsed pumps are used, cross phase modulation from the individual pumps causes broadening of the signal frequency channels at higher powers. Similarly, for $N=4$, as shown in Fig.~\ref{fig4}(c), an image can be seen at a larger pump power around 0.4mW, and there is a clear displaced image at 0.2mW, which corresponds to the point where the $k=0, \pi$ components have come back into phase, but the $k=\pm \pi/2$ components are out of phase by $\pi$.

Before forming an image in the Talbot effect, a displaced image can occur, when the relative phase-shifts between $k$ components are equal to an integer multiple of $k$. Here, a displacement of one lattice site corresponds to multiplication by a factor $\exp({ik})$ in reciprocal space. For the cases $N=2,4$ these displaced images can already be seen in Fig.~\ref{fig4}, occurring half way between the real images. However for $N=3$, it is necessary to introduce asymmetry to the band-structure (i.e. TRS breaking) to obtain a displaced image. Fig.~\ref{fig5}(a) shows the $N=3$ case, but with a $\pi/2$ phase-shift applied between the two pump frequencies. It can be seen that this creates an asymmetric band-structure where the $k$ components for $N=3$ lie along a straight line. 
Then, the components of the state accumulate phase differences proportional to $k$, and hence can form an image displaced by 1 site at 0.15mW pump power, then by 2 sites at 0.3mW.

Previously, the discrete Talbot effect has not been demonstrated with $N=5$ or $N\ge 7$, because  when there is only nearest-neighbor hopping these cases 
cannot satisfy the requirement of Eq.~(\ref{eq1})~\cite{Iwanow2005}. Here, we show that by inducing non-local hopping, this restriction can be lifted. We find that for $N=5$, the Talbot effect can be achieved by using three equally-spaced pumps, $A_1 = A_3 = 2 A_2$. This induces nearest-neighbor and next-nearest-neighbor hopping at equal rates, such that $\beta(k)\propto [\cos(k)+\cos(2k)]$, and 
\begin{equation}
\beta_m= (\beta_0/4) \left\{-1, -1, 4, -1, -1\right\},
\end{equation}
so an image can be formed. The experimental result is shown in Fig.~\ref{fig5}(b), and an image of the input signal is clearly seen at 0.17mW of pump power. This approach could be extended to larger $N$, with long-range hopping enabling flexible band-structure  engineering such that an image appears.

\section{Discussion and Conclusions}

In future work, it should be possible to apply spectral photonic lattices to single photons and correlated photons, so as to observe quantum interference effects and entanglement. 
Currently in our experiment the Raman gain of the nonlinear fiber generates a significant noise level which limits single photon experiments; this could be circumvented by moving the signal much further from the pump in wavelength while taking care to maintain phase-matching, by cooling the fiber to cryogenic temperatures as in ~\cite{Clemmen2016}, or making use of another nonlinear medium with more favourable properties.

In summary, we have demonstrated discrete spectral lattices with tunable hopping Hamiltonians enabled by the four-wave mixing Bragg scattering process. These can include non-local hopping and complex-valued coupling coefficients, by controlling the frequency separations and phases of the pump lasers. Complex-valued coupling coefficients imply a momentum-direction-dependent phase-shift which breaks time-reversal symmetry. The ability to tune multiple complex coupling coefficients corresponding to different lattice separations provides considerable freedom to engineer the band-structure. Besides, as the nonlinear frequency conversion only takes place one-way due to the phase-matching condition, this pumped system is non-reciprocal~\cite{Wang2017} and may be considered in applications. 

\section{Funding}

Australian Research Council: Centre of Excellence CUDOS (CE110001018); Laureate Fellowship (FL120100029); Discovery Project (DP160100619).


\section{Appendix 1: Equations of motion and dispersion relations} \label{sect:equations}

We begin by assuming that the signal and pump wavefunctions, $a(t,z)$ and $A(t,z)$ respectively, consist of a lattice of discrete frequencies with complex amplitudes $a_n(z)$ and $A_n(z)$ 
\begin{equation}
a(t,z)=\sum_{n=-\infty}^{+\infty} a_n(z) e^{-i\Omega nt}, \quad A(t,z)=\sum_{n=-\infty}^{+\infty} A_n(z) e^{-i\Omega nt},
\end{equation}
where $\Omega$ is the channel spacing in angular frequency and $z$ the position along the fiber. The wavefunctions evolve with $z$ according to nonlinear Schr\"{o}dinger equations (NLSE), with effective nonlinearity $\gamma$,
\begin{equation}
\partial_za(t,z)=2i\gamma |A(t,z)|^2 a(t,z), \quad \partial_zA(t,z)=i\gamma |A(t,z)|^2 A(t,z),
\end{equation}
where the signal and pump move at the same group-velocity and dispersion can be ignored - in practice this is achieved by placing them equidistant in frequency to either side of a zero dispersion wavelength. The signal is also weak enough that the pump evolution is independent of it; hence the pump experiences a nonlinear self-phase-shift, but its intensity as a function of time is unchanged:
\begin{equation}
|A(t,z)|^2=|A(t,0)|^2=\sum_{m=-\infty}^{+\infty} \sum_{l=-\infty}^{+\infty} A_m(0) A_l^\ast(0) e^{i\Omega (l-m)t}.
\end{equation}
Inserting this expression into the signal's NLSE, we obtain
\begin{equation}
\partial_za(t,z)=2i\gamma\sum_{m=-\infty}^{+\infty} \sum_{l=-\infty}^{+\infty} A_m(0) A_l^\ast(0) e^{i\Omega(l-m)t}~a(t,z).
\end{equation}
Then, we expand $a(t,z)$ into frequency components and group the frequencies together, and get equations of motion for the individual frequency components $a_n$
\begin{equation}
\partial_za_n(z)=2i\gamma\sum_{m=-\infty}^{+\infty} \sum_{l=-\infty}^{+\infty} A_m(0)A_l^\ast(0)~a_{n+m-l},
\end{equation}
i.e. each pair of pump frequency components, $A_m$ and $A_l$, creates coupling between signal channels separated by $m-l$.

The equations of motion can also be written in terms of $j$th-order coupling coefficients $c_j$ by grouping together pairs of pumps which are separated by $j$:
\begin{equation}
\partial_za_n=i \sum_{j=-\infty}^{+\infty} c_j a_{n+j}
             = i\sum_{j=1}^{+\infty} \left[ c_j a_{n+j}+c^\ast_j a_{n-j} \right] + i c_0 a_n, 
            \quad c_j=2\gamma\sum_{m=-\infty}^{+\infty}A_m(0) A_{m-j}^\ast(0) .
\end{equation}
Note that the coupling coefficients are symmetric, $c_j \equiv c_{-j}^\ast$, and accordingly the evolution is conservative, $\partial_z \sum_n |a_n|^2 \equiv 0$.
The term $c_0$ does not result in hopping but rather contains the cross phase-shifts from individual pump frequencies to the signal. This does not affect the dynamics and can be removed via a change of reference frame, $a_n \rightarrow a_n \exp(i c_0 z)$, which maps $c_0 \rightarrow 0$. We assume such a transformation in our analysis.

The hopping provides an effective tight-binding potential which is periodic in frequency every $\Omega$, and accordingly a dispersion relation or band-structure can be used to describe the evolution, where time plays the role of wave-vector. We use a notation $k=\Omega t$ to emphasize that this quantity plays the role of wave-vector, rather than having the state evolving with time. Then, a signal spectrum $a_n(z)=\exp\{i[kn+\beta(k)z]\}$ is a steady state solution with eigenvalue:
\begin{equation}
\begin{split}
\beta(k) &=\sum_{j=1}^{+\infty} \left[ c_j e^{ijk}+c_j^\ast e^{-ijk} \right] \\
        &= 2\gamma\sum_{m=-\infty}^{+\infty} \sum_{l=-\infty}^{m-1} 
             \left[ A_m(0) A^\ast_l(0) e^{i(m-l)k}+A_m^\ast(0) A_l(0) e^{-i(m-l)k} \right] \\
         &= 2\gamma\sum_{m\neq l} A_m(0) A^\ast_l(0) e^{i(m-l)k} .
\end{split}
\end{equation}
%

\section{Appendix 2: Simulations}  \label{sect:simulations}

Here, we provide simulations in comparison with the experimental results from the main paper. The simulations made use of the fast-fourier transform (FFT) function in Matlab to solve the NLSE for the signal - the input signal and pump spectra were transformed into the time-domain, then the time-dependent cross-phase-shift experienced by the signal was applied. Finally the signal was transformed back into the frequency domain to obtain the output spectrum. This approach assumes that dispersion can be ignored, as described above, but can include the finite bandwidths of the pumps so that some of the resulting broadening of the individual signal channels can be seen. Simulated results without this broadening are also shown.

\begin{figure}[b]
	\centering
	\includegraphics[width=\textwidth]{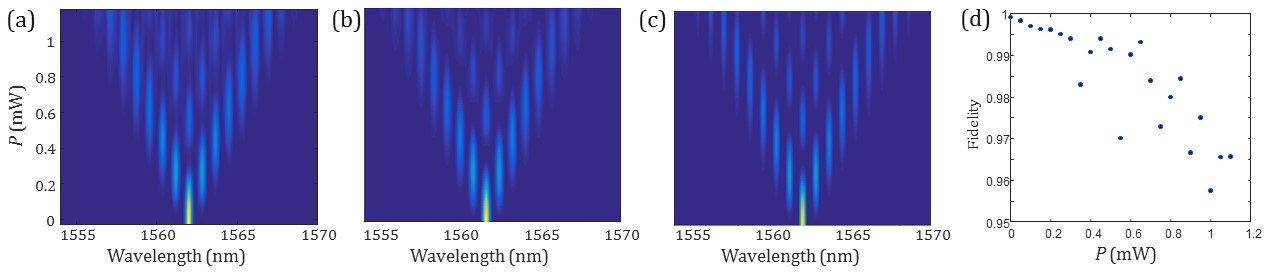}
        \vspace{-5pt}
	\caption{(a) Experimental result, (b) simulated result with realistic pump bandwidths, and (c) ideal case for diffraction of a single-site input under the influence of two pumps with a frequency separation of 100GHz.(d) Fidelity between measured and ideal spectrum as a function of pump power.
}\label{figS1}
 \vspace{-5pt}
\end{figure}

Figure~\ref{figS1}(a) shows the measured result for the simple case of two pumps separated by 100GHz and excitation of a single input site in the signal spectrum. Fig.~\ref{figS1}(b,c) shows the corresponding simulations, where the range of pump powers has been tuned to match the experimental result. The two contain very similar diffraction patterns as expected. Fig.~\ref{figS1}(d) shows the fidelity between the measured spectrum and an ideal spectrum as a function of power. This ideal spectrum is taken from a simulation with narrow-band pumps, so that there is no broadening of the individual channels. The signal spectrum is divided up into 100GHz broad channels and the total power within each channel is found, then each spectrum is normalised to sum to 1. The fidelity $F$ is calculated according to $F=\sum_n |a_n^{exp}||a_n^{ideal}|$ where the absolute values of the experimental amplitudes $|a_n^{exp}|$ are the square-roots of the intensities. This fidelity decreases slightly and fluctuates at higher pump power, because there the experimental spectra are affected by factors such as broadening from the finite-bandwidth pumps, the third-order dispersion of the fibre, and the introduction of Raman scattered noise from the pumps. However the fidelity remains $>95\%$ over the range of pump powers used.

\begin{figure}[t]
	\centering
	\includegraphics[width=.85\textwidth]{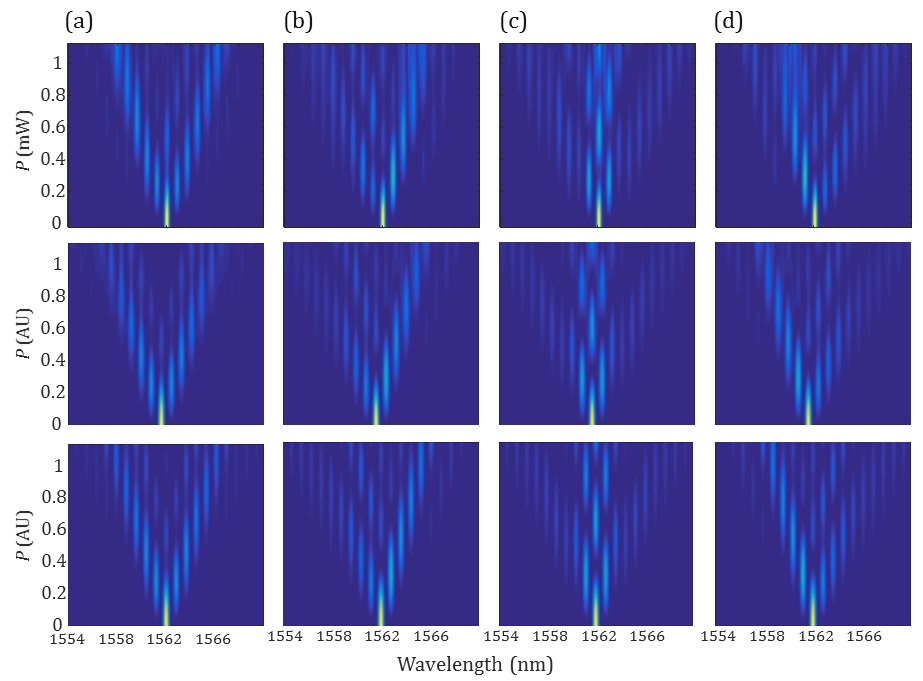}
        \vspace{-5pt}
	\caption{Experimental (top row) and simulation (middle and bottom rows) results when a third pump is used to introduce 2nd and 3rd order coupling. The middle row includes the broadening due to finite pump bandwidths, the bottom row is the ideal case. The phase of the third pump is set to (a) 0, (b) $\pi/2$, (c) $\pi$, and (d) $3\pi/2$ radians.
}\label{figS2}
 \vspace{-5pt}
\end{figure}

Figure~\ref{figS2} shows the four cases where a weaker third pump is added to introduce 2nd and 3rd order hopping, with the phase of the third pump shifted by 0, $\pi/2$, $\pi$, and $3\pi/2$ radians. The top row shows the experimental results and the bottom row the simulations. In each case qualitatively very similar behavior is observed, though at higher pump powers some discrepancies can be seen. Partly this is because of the broadening of individual signal channels due to the finite bandwidths of the pump channels. Imperfections could also result from the small third-order dispersion of the nonlinear fiber, or from phase errors between the three pumps, originating from the original mode-locked laser pulse, the amplifier, or the spectral pulse shapers.

\begin{figure}[h]
	\centering
	\includegraphics[width=.75\textwidth]{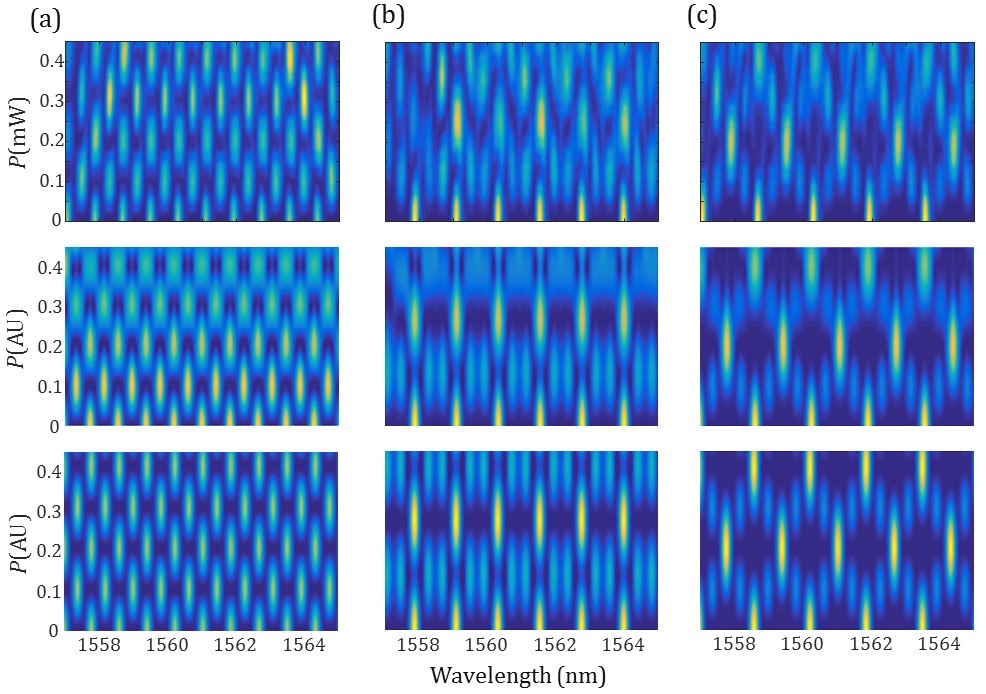}
        \vspace{-5pt}
	\caption{Experimental (top row) and simulation (middle and bottom rows) results for the spectral discrete Talbot effect with periodicities (a) $N=2$, (b) $N=3$, and (c) $N=4$. Two pump frequencies are used without phase-shifts. The middle row includes the broadening due to the finite pump bandwidths, the bottom row is the ideal case.
}\label{figS3}
 \vspace{-5pt}
\end{figure}

Figure~\ref{figS3} shows the results for the spectral discrete Talbot effect using only 1st order coupling (two pumps) with a real coupling coefficient, with input signals which are periodic every 2, 3, or 4 lattice sites. The experimental results (top row) show the formation of images as predicted by the simulations (bottom row), although the experimental images show varying degrees of distortion; in particular, some occupied sites have higher intensities than others, and some sites which should not be occupied contain some background. Since these measurements make use of a large number of input sites for the signal, spread over a wide bandwidth, phase-errors between input sites are an additional source of experimental imperfection. The first spectral pulse shaper in the setup was used to attempt to correct the phase of each input site, relative to its neighboring sites, but some residual errors are likely to have been present.

\begin{figure}[h]
	\centering
	\includegraphics[width=.7\textwidth]{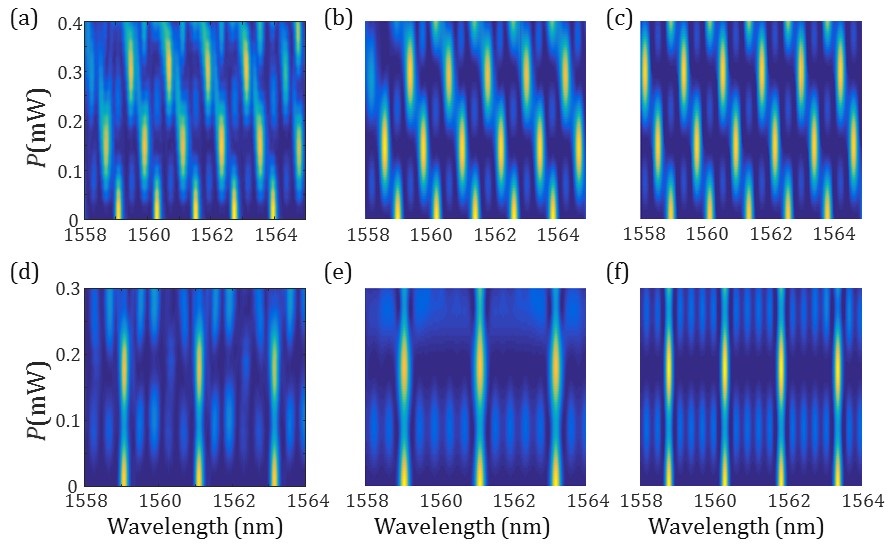}
        \vspace{-5pt}
	\caption{(a) Experimental result, (b) simulated result with finite pump bandwidths, and (c) simulated ideal case for the $N=3$ discrete Talbot effect with an imaginary coupling coefficient to break time-reversal symmetry. (d) Experimental result, (e) simulated result with finite pump bandwidths, and (f) simulated ideal case for the $N=5$ discrete Talbot effect, requiring a balance of 1st and 2nd order hopping.
}\label{figS4}
 \vspace{-5pt}
\end{figure}

Figure~\ref{figS4}(a) and (b) show the experimental and simulated result for the $N=3$ discrete Talbot effect when a $\pi/2$ phase-shift is applied to one of the pumps, resulting in an imaginary hopping coefficient between neighboring waveguides. The propagation is asymmetric and displaced images appear - good agreement is seen between experiment and simulation.

Figure~\ref{figS4}(c) and (d) show the experimental and simulated result for the $N=5$ discrete Talbot effect, which requires a 2nd order coupling coefficient introduced by a third pump in order for an image to form. There is good agreement between experiment and simulation, with a clear image of the input state occurring.

\end{document}